\documentclass[12pt]{amsart}

\usepackage[hmargin=0.8in,twosideshift=0in,height=8.6in]{geometry}
\usepackage{amssymb,amsthm}
\usepackage{delarray}
\usepackage{natbib}

\usepackage{ifpdf}
\ifpdf
\usepackage[pdftex]{graphicx}
\else
\usepackage[dvips]{graphicx}
\fi

\usepackage{ifpdf}
\usepackage{color}
\definecolor{webgreen}{rgb}{0,.5,0}
\definecolor{webbrown}{rgb}{.8,0,0}
\definecolor{emphcolor}{rgb}{0.95,0.95,0.95}

\usepackage{hyperref}
\hypersetup{%
          colorlinks=true,
          linkcolor=webbrown,
          filecolor=webbrown,
          citecolor=webgreen,
          breaklinks=true}
\ifpdf
\hypersetup{pdftex,
            pdfstartview=FitH, 
            bookmarksopen=true,
            bookmarksnumbered=true
}
\else
\hypersetup{dvips}
\fi

\numberwithin{equation}{section} 

\newtheorem{prop}{Proposition}
\newtheorem{rem}{Remark}
\newtheorem{example}{Example}

\newtheorem{definition}{Definition}

{\bf}{\it}
\newtheorem{remark}{Remark}
{\bf}{\it}
\newtheorem{corollary}{Corollary}

\title[No Arbitrage Conditions For Simple Trading Strategies]{No Arbitrage Conditions For Simple Trading Strategies}
\author{Erhan Bayraktar}
\address[Erhan Bayraktar]{Department of
  Mathematics, University of Michigan, Ann Arbor, MI 48109}
\email{erhan@umich.edu}
\thanks{E. Bayraktar is supported in part by the National Science Foundation. }

\author{  Hasanjan Sayit}
\address[Hasanjan Sayit] {Department of Mathematics, Worcester Polytechnic Institute.}
\email{hs7@WPI.EDU}



\begin{document}
\begin{abstract} 
Strict local martingales may admit arbitrage opportunities with respect to the class of simple trading strategies. (Since there is no possibility of using doubling strategies in this framework, the losses are not assumed to be bounded from below.)
We show that for a class of non-negative strict local martingales, the strong Markov property implies the no arbitrage property with respect to the class of simple trading strategies. This result can be seen as a generalization of a similar result on three dimensional Bessel process in \cite{DS1994}. We also provide no arbitrage conditions for stochastic processes within the class of simple trading strategies with shortsale restriction. 
\end{abstract}
\maketitle

\section{Introduction}

We consider a market with a risky asset with price process $X$ and a risk-free asset with price process $B$ in the time horizon $[0, \infty)$. We will assume that  $B_t \equiv1$, which corresponds to taking the bond price as the num\'{e}raire. The price processes are defined on a filtered
probability space $(\Omega,\mathcal{F},P,\mathbb{F}=(\mathcal{F}_t)_{t \geq0})$ satisfying the 
``usual hypotheses'' (i.e., the filtration $\mathbb{F}$ is right continuous, and $\mathcal{F}_0$ contains all of
the $P$ null sets of $\mathcal{F}$). In this paper, we restrict our trading strategies to the following class of simple integrands:  

\begin{definition} 
The set of simple trading strategies with short sale restriction is given by
$S^0(\mathbb{F})=
\{
g_01_{\{0\}}+\sum_{j=1}^{n-1}g_j1_{(\tau_j,\tau_{j+1}]}:n\geq 2, 0\le
\tau_1\le \ldots \le \tau_n
$
where all of the $\tau_j$ are bounded $\mathbb{F}$-stopping times; $g_0$ is
a non-negative real number,and the $g_j$ are non-negative real valued
$\mathcal{F}_{\tau_i}$ measurable random variables\}. The set of simple trading
strategies is denoted by $S(\mathbb{F})$ and is
given by $S(\mathbb{F})=S^0(\mathbb{F})-S^0(\mathbb{F})$.
\end{definition}

Let $K^0=\{(H\cdot X)_{\infty}|H\in S^0(\mathbb{F})\}, 
K^s=\{(H\cdot X)_{\infty}|H\in S(\mathbb{F})\} $ denote the
outcomes of the corresponding trading strategies respectively for any
adapted price process $X$.

\begin{definition} \label{DEF}
We say $X$ satisfies the no arbitrage property with respect to 
$S^0(\mathbb{F}), S(\mathbb{F})$ separately if 
$K^0\cap L_0^{+}=\{0\}, K^s\cap L_0^{+}=\{0\}$
respectively. (Here $L_{0}^{+}$ is the collection of equivalence classes of non-negative measurable functions on the probability space $(\Omega,\mathcal{F},P)$.)
\end{definition}

Observe that
we do not assume that the losses are bounded from below, which is not necessary since the class of doubling strategies is not a subset of either $S(\mathbb{F})$ or $S^0(\mathbb{F})$. But as it was observed in \cite{DS1994} strict local martingales may admit arbitrage opportunities with respect to $\mathbb{S}(F)$. In Section 2, we give a necessary and sufficient conditions, which we baptize as ``condition $(*)$", for local martingales to admit no arbitrage with respect to $S(\mathbb{F})$. We give examples of strict local martingales that do/do not admit arbitrage with respect to $S(\mathbb{F})$. We also provide an alternative proof of Theorem 6 of \cite{DS95} as an application of our no arbitrage characterization of local martingales. We also show that condition $(*)$ is preserved under composition with non-decreasing functions and give an application of this result.

In Section 3, we derive a sufficient and necessary condition for no-arbitrage, and show that the no arbitrage property is preserved under composition with strictly increasing functions. As an application of this result we can construct processes that are not semi-martingales but do not admit arbitrage with respect to $S^0(\mathbb{F})$.
 
\section{No Arbitrage Conditions wrt $S(\mathbb{F})$}

In this section we provide a necessary and sufficient condition for the no arbitrage property of non-negative \emph{strict local
martingales} (i.e. not true martingales, see e.g. \cite{DS95}) with respect to the simple trading strategies $S(\mathbb{F})$. A typical example of a strict local martingale is the inverse process of three dimensional Bessel process, see \cite{DS1994}. Strict local martingales appear in a number of ways in applications and some of  their financial applications were discussed in \cite{DS1994, DS95}. Also see \cite{DH, HS2000, sin} for stock price models with stochastic volatility for which the most natural candidates for the pricing measures is an equivalent strict local martingale measure.

It is well known that local martingales satisfy the no arbitrage property when the admissible trading strategies are such that accumulated losses are bounded below \cite{DS94}. This condition on the losses is added to avoid doubling strategies. However, the collection of doubling strategies is not a subset the class of simple trading strategies 
$S(\mathbb{F})$. Therefore, we do not impose lower bound for the gain process in our definition of no arbitrage with simple trading strategies. As a result strict local martingales may admit arbitrage possibilities. Here we will give necessary and sufficient no arbitrage conditions for strict local martingales.

One of the advantages of working with $S(\mathbb{F})$ is that, the no arbitrage property of a process  $X$ with respect to $S(\mathbb{F})$ implies the no arbitrage property of the process $f(X)$ with respect to $S(\mathbb{F})$ as well, for any strictly increasing or 
strictly decreasing function $f$, see Corollary 5 in \cite{HPJ}. However the same is not true for more general admissible strategies. For example, three dimensional Bessel process admits arbitrage with respect to the class of general admissible
trading strategies, see \cite{DS95}, while its inverse process, a non-negative strict local martingale, does not. This has the undesirable effect of having num\'{e}raire dependency in the existence of arbitrage (see Section 4 of \cite{DS95}).

To state the main result of this section, we introduce a condition 
that is weaker than stickiness (see e.g. \cite{BS}, \cite{Gua}). We will show that this condition is a necessary and sufficient condition
for the no arbitrage property with respect to $S(\mathbb{F})$ of non-negative strict local martingales. This characterization will allow us to state a general result on non-negative local martingales with strong Markov property.

\begin{definition}\label{defn:ltstick} We say that  an adapted  c\`adl\`ag process $X$ satisfies condition $(\star)$ with respect
  to the filtration $\mathbb{F}$ if for any bounded stopping time
  $\tau$ and any $A\in \mathcal{F}_{\tau}$ with $P(A)>0$ we have
\[
P\left(A\cap \left\{\inf_{t\in [\tau, T]}(X_t-X_{\tau})> -\epsilon \right\}\right)>0
\]
for any $\epsilon>0$ and $T$ with $\tau\le T$ a.s. 
\end{definition}

\begin{remark} \label{remark1} The condition $(\star)$ is a weaker
  condition than the sticky condition, compare Definition 2.9 of \cite{Gua} with Definition~\ref{defn:ltstick}. In
  \cite{Gua}, it was shown that strong Markov processes with 
  regular points and continuous processes with full support on the 
space of continuous functions are sticky. As a result all these processes satisfy 
condition $(\star)$. 
\end{remark}
We are now ready to state the main result of this section:
\begin{prop}\label{main} Assume $X$ is a nonnegative 
c\`adl\`ag $\mathbb{F}$-semimartingale and $X$ admits an equivalent local martingale measure $Q$. 
Then, $X$ satisfies the no arbitrage property in $S(\mathbb{F})$ if and only if 
$X$ satisfies the condition $(\star)$.
\end{prop}
\begin{proof} \emph{Sufficiency}: Assume $X$ satisfies the condition $(\star)$ and that there exists arbitrage opportunities. Since $X$ is bounded below and 
  admits an equivalent local martingale measure, it is a
  supermartingale under this equivalent measure. Without loss of generality we can assume its arbitrage strategy is of the form $-1_{(\tau_0, \tau_1]}$ for two bounded stopping times $\tau_0 \le
\tau_1$ (see Lemma 5 of \cite{HPJ}). So we assume  $X_{\tau_0}\geq
X_{\tau_1}$ a.s. and $P(X_{\tau_0}>X_{\tau_1})>0$. Let $K$ be a number
such that the event $A=\{X_{\tau_0}<K\}\cap \{\tau_1>\tau_0\}$ satisfies
$P(A\cap \{X_{\tau_0}>X_{\tau_1}\})>0$. Note that there is such a $K$
because the event $\{X_{\tau_0}>X_{\tau_1}\}$ has positive probability
and  $\{X_{\tau_0}>X_{\tau_1}\}\subset \{\tau_1>\tau_0\}$ and that $X$ is a c\'{a}dl\'{a}g super-martingale, which implies that $X$ is a.s. bounded on a compact domain see e.g. Theorem 1.3.8 of \cite{KS1991}. Also we
have that $A\in \mathcal{F}_{\tau_0}$. Fix
any number $T$ with $T\geq \tau_1$ a.s. and
define the following two stopping times
\[
\tau_0^A=
\left\{
\begin{array}{ll}
\tau_0 & \mbox{if $\omega \in A$}, \\
  T    & \mbox{if $\omega \notin A$}
\end{array}
 \right.
\]
and
\[
\tau_1^A=
\left\{
\begin{array}{ll}
\tau_1 & \mbox{if $\omega \in A$}, \\
  T    & \mbox{if $\omega \notin A$}
\end{array}
 \right..
\]

Let $\tau=\inf\{t\geq \tau_0^A: X_t>K+1 \}\wedge \tau_1^A$. In what follows we use $N$ to denote sets of measure zero. (Although these sets might differ we use the same letter to avoid notational crowding.)
If $\tau=\tau_1^A$ on $A-N$ then since $X_{\tau_0}\le K$ and
$X_{\tau_1}\le K$ on $A-N$ we have that the local martingale 
$1_A(X_t-X_{\tau_0})$ is bounded in $[\tau_0, \tau_1]$. So it is a
martingale in $[\tau_0, \tau_1]$. This contradicts with
$P(A\cap \{X_{\tau_0}>X_{\tau_1}\})>0$. So we assume the event 
$B=A\cap \{\tau<\tau_1^A\}$ has positive probability. Note that 
$B\in \mathcal{F}_{\tau}$. Since $X$
is c\`adl\`ag, $X_{\tau}\geq K+1$  on $B-N$. Since $X_{\tau_1}\le K$
on A-N we have that
$P(B\cap \{\inf_{t\in [\tau,
  T]}(X_t-X_{\tau})>-\frac{1}{2}\})=0$. This contradicts with
condition $(\star)$. So $X$ satisfies the no arbitrage property in $S(\mathbb{F})$. 

\emph{Necessity:}
Assume $X$ has no arbitrage strategy in 
$S(\mathbb{F})$ and $X$ does not satisfy condition $(\star)$. Then there
is a bounded stopping time $\tau$ and $A\in
\mathcal{F}_{\tau}$ with $P(A)>0$ and  $T, \epsilon>0$ with $T\geq \tau$
a.s. such that $P(A\cap \{\inf_{t\in [\tau,
T]}(X_t-X_{\tau})>-\epsilon\})=0$. So $\inf_{t\in [\tau,
T]}(X_t-X_{\tau})\le -\epsilon $ on $A/N$ for a measure zero set $N$.  
Let $\tau^A=\tau$ on $A$ and 
$\tau^A=T$ on the complement of $A$.  Then $\tau^A$ is a stopping
time. If we define 
$\theta=\inf\{t\geq \tau^A: (X_t-X_{\tau^A})< -\frac{\epsilon}{2}\}$, 
then since $X$ has right continuous paths
$(X_{\theta}-X_{\tau^A})\le -\frac{\epsilon}{2}$ on $A/N$ and also we
have $\theta \le T$ on $A/N$. Let 
$\theta^A=\theta$ on $A$ and $\theta^A=T$ on the complement of $A$,
then $-1_{(\tau^A, \theta^A]}$ is an arbitrage strategy for $X$ in
$S(\mathbb{F})$. This completes the proof.\hfill 
\end{proof}
\begin{remark}
We should mention that Proposition~\ref{main}  is useful only for strict local 
martingales. This is because all martingales satisfy condition $(\star)$.
\end{remark}
\begin{remark} \label{remarkk} We remark that the second part of the proof of the above proposition shows that
if $X$ satisfies the no arbitrage property in $S(\mathbb{F})$, then $X$ satisfies the condition $(\star)$. This fact will be used in the proof of Corollary \ref{exp} below.
\end{remark}
\begin{remark} We remark that the condition in Proposition \ref{main}  that $X$ is non-negative can be relaxed. We can replace this condition by the requirement that  the negative part of $X$  belongs to class DL, a class of processes $X$ such that for each $a>0$, $\{X_S\}$ is uniformly integrable over all bounded stopping times $S\le a$. We can do this replacement thanks to Proposition 2.2 of \cite{KXM},
which shows that the local martingales whose negative part are of class DL are supermartingales.
\end{remark}
An immediate corollary of Proposition~\ref{main} is the following general result
on nonnegative semimartingales with strong Markov property.

\begin{corollary} \label{local} Let $X$ be a non-negative c\'{a}dl\'{a}g $\mathbb{F}$-semimartingale that admits  an equivalent local martingale measure. If $X$ is sticky, then it satisfies the no arbitrage 
property in $S(\mathbb{F})$. Especially if $X$ has strong Markov property (under the original measure) and if all points of $X$ are regular, then $X$ satisfies the no arbitrage property with respect to 
$S(\mathbb{F})$. 
\end{corollary}

\begin{proof} If $S$ is sticky, then it satisfies the condition $(\star)$ and Proposition \ref{main} applies   (compare Definition~\ref{defn:ltstick} with Definition 2 of \cite{BS} and also see Proposition 1 of \cite{BS}). If $X$ has strong Markov property and all of its points are regular, then it is sticky (see, \cite{Gua}) and again Proposition \ref{main} applies.    
\hfill  \end{proof}

Note that not all non-negative local martingales satisfy the condition
$(\star)$, the following is an example, given in \cite{DS1994},  of a 
non-negative process that admits an equivalent local martingale measure, but 
has arbitrage in
$S(\mathbb{F})$. 
\begin{example} Let
  $(B_t)_{t\geq 0}$ be a Brownian motion with $B_0=1$. Let
  $\tau=\inf\{t>0: B_t=0\}$. It is well know that $\tau$ is
  a.s. finite. Let 
\[
S_t=\left \{ 
\begin{array}{ll} 
B_{tan(\frac{t\pi}{2})\wedge \tau} & \mbox{when $t<1$}, \\ 
B_{\tau}=0 & \mbox{when $t=1$}.
\end{array}
\right.
\] Then $S$ is a nonnegative process that admits an equivalent local
martingale measure, see e.g. \cite{DS1994}.
The arbitrage strategy, $H$, of the process $S$ is given by $H=-1$ on
$(0, 1]$. Clearly $(H\cdot S)_1=1$ a.s. and $(H\cdot S)_0=0$ a.s. 
Observe that $S$ does not satisfy the condition $(\star)$: we can let 
$A=\Omega, \tau=0, T=1, \epsilon =0.5$ and obtain
$P(\inf_{t\in [0, 1]}\{t: (S_t-1)>0.5 \})=0$.  
\end{example}
A typical example of a non-negative local martingale that is not true 
martingale and that has strong Markov property is the inverse process of 
three dimentional Bessel process. 
The following is another example of a strict local martingale that satisfies the no arbitrage property in $S(\mathbb{F})$. 
\begin{example} Consider the CEV model
\[
dX_t=a X_tdt+bX_t^{\rho}dB_t.
\]
It is well known that the CEV models admit an equivalent martingale measure when $\rho\in (0, 1]$ and they admit a strict local martingale measure when $\rho >1$, see e.g. \cite{DH}.
CEV models are regular strong Markov processes: for the case $\rho>1$ this is stated in \cite{HS2000}, on the other hand for $\rho \in (0,1)$ this fact follows from Sections 5.4 and 5.5 of \cite{KS1991} (since the conditions ND' and LI' on p 343 hold). Finally observe that $\rho=1$ is the classical geometric Brownian motion model. 
As a result CEV processes are sticky and they satisfy condition $(\star)$ (see Corollary~\ref{exp}) and therefore do not admit arbitrage possibilities with respect to $S(\mathbb{F})$ by Proposition~\ref{main}.

In fact \cite{DH} considered a more general class of processes of the form
\[
dX_t=\sigma(X_t)dW_t, 
\]
and gave necessary and sufficient conditions on the function $\sigma$ for these processes to be martingales/strict local martingales (see Theorem 1.6 of \cite{DH}). By assuming that $\sigma$ also satisfies the ND' and LI' conditions mentioned above we obtain a large class of strict local martingales that admit no arbitrage with respect to $S(\mathbb{F})$.

\end{example}

The following corollary is the content of the Theorem 6 of \cite{DS95}. Here we restate it and show a simple proof as an application of Proposition \ref{main}.

\begin{corollary} \label{bessel} Let $X$ be the Bessel process of dimension
  $\delta\in (2, \infty)$ (see \cite{RY} Definition 11.1.1) and $X_0>0$. Then $X$ satisfies the no
  arbitrage property in $S(\mathbb{F})$.
\end{corollary}

\begin{proof} The process $M_t=X^{2-\delta}$ is a nonnegative
  local martingale (see \cite{RY} Chapter 11, Exercise 1.16) and a regular strong
  Markov process. By Proposition 2 of  \cite{BS} $M$ has the sticky
  property and so it satisfies the condition $(\star)$. Now applying Proposition 
\ref{main}, we see that $M$ satisfies the no arbitrage property with respect to
$S(\mathbb{F})$. Since $X$ is a composition of $M$ with strictly
decreasing function $f(x)=x^{-\frac{1}{\delta -2}}$ on $(0, \infty)$
, from Corollary 5 in \cite{HPJ} the result follows.  
\hfill  \end{proof}

The following proposition shows that the property 
$(\star)$ is invariant under composition with a
continuous nondecreasing function.

\begin{prop}\label{starmonotone} Let $X$ be a stochastic process
  adapted to the filtration
$\mathbb{F}$ and takes values in the interval $(a,b)$(including the cases when
$a=-\infty$ and/or $b=\infty$). If $X$
satisfies condition $(\star)$, then for any continuous
nondecreasing function $f$ defined on
$(a,b)$, the process $f(X)$ also satisfies  condition $(\star)$.
\end{prop}
\begin{proof} First we assume $a, b$ are bounded. We need to show for
  any bounded stopping time $\tau$ and
  any $A\in \mathcal{F}_{\tau}$ with $P(A)>0$, we have 
$P(A\cap \{\inf_{t\in [\tau, T]}(f(X_t)-f(X_{\tau}))> -\epsilon\})>0$
for any $\epsilon >0$ and $T$ with $\tau\le T$ a.s. Since $X_{\tau}$
takes values on $(a, b)$ and $P(A)>0$, for sufficiently large 
$n_0\in \mathbb{N}$, the
event $B=A\cap \{X_{\tau}\in [a+\frac{1}{n_0}, b-\frac{1}{n_0}]\}\in \mathcal{F}_{\tau}$ has
positive probability. The function $f$ is continuous
on $(a,b)$, therefore it is uniformly continuous on 
$[a+\frac{1}{n_0}, b-\frac{1}{n_0}]$.
That is for a given $\epsilon>0$,  there exists $\delta>0$ such that
whenever  $|x-y|<\delta$ and $x,y \in [a+\frac{1}{n_0},
b-\frac{1}{n_0}]$ we have $|f(x)-f(y)|<\epsilon$.
Since $X$ satisfies condition $(*)$ we have that
$P(B\cap \{\inf_{t\in [\tau, T]}(X_t-X_{\tau})>- \delta\})>0$. Since $f$ is uniformly continuous and non-decreasing  we have $B\cap \{\inf_{t\in [\tau, T]}(X_t-X_{\tau})>-
\delta\}\subset B\cap \{\inf_{t\in [\tau, T]}(f(X_t)-f(X_{\tau}))> -\epsilon\}$.
So $P(A\cap \{\inf_{t\in [\tau, T]}(f(X_t)-f(X_{\tau}))>
-\epsilon\})>0$ since $B \subset A$. When $a=-\infty$ and/or $b=\infty$, the above proof
can be adjusted by replacing $a+1/n_0$ with $a_n\downarrow -\infty$ and/or 
$b-1/n_0$ with $b_n\uparrow \infty$.   
\hfill  \end{proof}
 
 An application of Proposition~\ref{starmonotone} is the following result:
 
\begin{corollary}\label{exp} Let $X$ be a continuous semimartingale and assume $X$ admits 
an equivalent local martingale measure $Q$. Then $X_t-\frac{1}{2}[X, X]_t$
admits no arbitrage with respect to $S(\mathbb{F})$ if and only if 
$X_t-\frac{1}{2}[X, X]_t$ satisfies condition $(\star)$. 
\end{corollary}
\begin{proof} If $X_t-\frac{1}{2}[X, X]_t$ has no arbitrage in 
$S(\mathbb{F})$ then $X_t-\frac{1}{2}[X, X]_t$ satisfies  the condition $(\star)$ 
 (see Remark \ref{remarkk} above). Conversely, if $X_t-\frac{1}{2}[X, X]_t$ satisfies the condition 
$(\star)$ then, by Proposition \ref{starmonotone} above, $e^{X_t-\frac{1}{2}[X, X]_t}$ also satisfies the condition $(\star)$. Since $e^{X_t-\frac{1}{2}[X, X]_t}$ is a local martingale under the measure $Q$,  
by Proposition \ref{main} it satisfies no arbitrage with respect to
$S(\mathbb{F})$. Therefore by Corollary 5 of \cite{HPJ},  
$X_t-\frac{1}{2}[X, X]_t$ also satisfies no arbitrage with respect to $S(\mathbb{F})$.
\hfill  \end{proof}

\section{No Arbitrage with Shortsales Restrictions}
 Let us denote 
$L_{++}^0(\Omega,\mathcal{F},\mathbb{P}):=\{\eta \in L^0(\Omega,\mathcal{F},\mathbb{P}): P(\eta \geq 0)=1\;\; \mbox{and} \;\;
P(\eta > 0)>0 \}$.

\medskip

\noindent We begin with the characterization of no arbitrage with respect to $S^0(\mathbb{F})$.

\begin{prop}\label{key lemma}
An adapted c\'{a}dl\'{a}g process $X_t,t\in [0, \infty)$ satisfies the  no arbitrage property in
$S^0(\mathbb{F})$ if and only if for any two bounded stopping times
$\tau_1\geq \tau_0$, and any $A\in \mathcal{F}_{\tau_0}$ we have
$1_A(X_{\tau_1}-X_{\tau_0})\in L^0(\Omega,\mathcal{F},\mathbb{P})\setminus L_{++}^0(\Omega,\mathcal{F},\mathbb{P})$.
\end{prop}

\begin{rem}
We have $\xi \in L^0(\Omega ,\mathbb{F} ,P)
\setminus L_{++}^0(\Omega, \mathbb{F}, P)$ 
if and only if $\xi \in L^0(\Omega ,\mathcal{F} ,Q)
\setminus L_{++}^0(\Omega, \mathcal{F}, Q)$  when $Q$
is equivalent to $P$. 
\end{rem}

\noindent\textbf{Proof of Proposition~\ref{key lemma}}  \emph{Necessary condition for no arbitrage.}
If we assume that $1_A(X_{\tau_1}-X_{\tau_0})\in L^0_{++}$ for two bounded stopping
times $\tau_0\le \tau_1$ and $A\in \mathcal{F}_{\tau_0}$, then  
$1_A1_{(\tau_0, \tau_1]}\in S^0(\mathbb{F})$ is an arbitrage strategy
for $X$ (i.e. no arbitrage implies $1_A(X_{\tau_1}-X_{\tau_0})\in L^0\setminus L_{++}^0$).

\emph{Sufficient condition for no arbitrage.} Assume that for any two bounded stopping times
$\tau_0\le \tau_1$ and any $A\in \mathcal{F}_{\tau_0}$ we have 
$1_A(X_{\tau_1}-X_{\tau_0})\in L^0\setminus L_{++}^0$ and that $X$ admits 
arbitrage. Assume that the arbitrage strategy is given by 
$V=g_01_{\{0\}}+\sum_{j=1}^{n-1}g_j1_{(\tau_j,\tau_{j+1}]}\in S^0(\mathbb{F})$ with
$P(g_j>0)>0$ for some $j\in \{1,2,\ldots,n-1\}$ such that
$(V\cdot X)_T\geq 0$ a.s and $P((V\cdot X)_T>0)>0$. Let
\begin{eqnarray*}
k &=& \min\bigg\{l \in \{0, \cdots, n-1\}:P(g_l>0)>0,\;\;
P(\sum_{j=1}^lg_j(X_{\tau_{j+1}}-X_{\tau_j})\geq 0)=1,\\
& & {} P(\sum_{j=1}^lg_j(X_{\tau_{j+1}}-X_{\tau_j})>0)>0 \bigg\}.
\end{eqnarray*}
(Note that $k$ is well-defined because we assumed that the arbitrage strategy is given by $V$.)
If $k=1$ then $P(g_1>0)>0$ and $g_1(X_{\tau_2}-X_{\tau_1})\geq 0$ a.s and
 $g_1(X_{\tau_2}-X_{\tau_1})> 0$ with positive probability. Let
 $C=\{g_1>0\} \in \mathcal{F}_{\tau_1}$. Since 
$g_1(X_{\tau_2}-X_{\tau_1})\geq 0 \;\;
 \mbox{a.s.}$ we have $X_{\tau_2}\geq X_{\tau_1}$ on $C/N$ in which $N$ is a null set. Because  
$g_1(X_{\tau_2}-X_{\tau_1})> 0$ with positive probability we have  that
$X_{\tau_2}>X_{\tau_1}$ with positive probability on $C$. 
As a result $1_C1_{(\tau_1, \tau_2]}\in L_{++}^0$ which contradicts our assumption. So we
assume $k>1$. From the definition of $k$, we either have 
$\sum_{j=1}^{k-1}g_j(X_{\tau_{j+1}}-X_{\tau_j})=0$ a.s. or 
$\sum_{j=1}^{k-1}g_j(X_{\tau_{j+1}}-X_{\tau_j})<0$ with positive
probability. If $\sum_{j=1}^{k-1}g_j(X_{\tau_{j+1}}-X_{\tau_j})=0$
a.s.,  then $g_k(X_{\tau_{k+1}}-X_{\tau_k})\geq 0$ a.s. and 
$P(g_k(X_{\tau_{k+1}}-X_{\tau_k})>0)>0$. If we let $C=\{g_k>0\}$,
we have $1_{C}1_{(\tau_{k},\tau_{\tau_{k+1}}]}\in L_{++}^0$ which again contradicts our assumption. Let us assume
$\sum_{j=1}^{k-1}g_j(X_{\tau_{j+1}}-X_{\tau_j})<0$ with positive probability. Let
$C=\{\sum_{j=1}^{k-1}g_j(X_{\tau_{j+1}}-X_{\tau_j})<0\}$, then 
$C\in \mathcal{F}_{\tau_{k-1}}$ and 
$P(C)>0$ and since $\sum_{j=1}^{k}g_j(X_{\tau_{j+1}}-X_{\tau_j})\geq
0$ a.s. we have $g_k(X_{\tau_{k+1}}-X_{\tau_k})>0$ on
$C$. Since $g_k\geq 0$ a.s., we have $X_{\tau_{k+1}}>X_{\tau_k}$ on
$C$, which implies that $1_C(X_{\tau_{k+1}}-X_{\tau_k})\in L_{++}^0$. This contradicts our assumption. We can now conclude that X admits no arbitrage with respect to $S^0(\mathbb{F})$. \hfill


The next result the no arbitrage property with respect to $S^0(\mathbb{F})$ is closed under composition with strictly increasing functions.

\begin{prop}\label{monotone1}
Let $X=(X_t)_{t\geq 0}$ be a c\`adl\`ag stochastic process adapted to the
filtration $\mathbb{F}$ and let $f$ be any strictly increasing
continuous function whose domain contains the
range of $X$. Then the no arbitrage property of $X$ in $S^0(\mathbb{F})$,
is equivalent to the no arbitrage property of $Y_t=f(X_t)$ in
$S^0(\mathbb{F})$.
\end{prop}
\begin{proof}
Assume $X_{t}$ satisfies the no arbitrage property in
$S^0(\mathbb{F})$. We need to show for 
any $\tau _{1}\geq \tau _{0}$ and any $A\in \mathcal{F} _{\tau _{0}}$
we have $1_{A}(f(X_{\tau _{1}})-f(X_{\tau _{0}}))\in
L^0\setminus L^0_{++}$ (Lemma \ref{key lemma} ). Since $X$ has no arbitrage in
$S^0(\mathbb{F})$, we have $1_A(X_{\tau_1}-X_{\tau_0})\in L^0\setminus
L_{++}^0$. This implies that either
$X_{\tau_1}=X_{\tau_0}$ a.s. on $A$ or $X_{\tau_1}<X_{\tau_0}$ with
positive probability on $A$. Since $f$ is strictly increasing, we have
either $f(X_{\tau_1})=f(X_{\tau_0})$ a.s. on $A$ or  
$f(X_{\tau_1})<f(X_{\tau_0})$ on $A$ with positive probability and
this implies $1_A(f(X_{\tau_1})-f(X_{\tau_0}))\in L^0\setminus
L_{++}^0$. As a result again by applying Proposition \ref{key lemma} we observe that $f(X)$ satisfies the no
arbitrage property with respect to $S^0(\mathbb{F})$. 

Now assume that $f(X_t)$ satisfies the no arbitrage property.
Since $X_{t}=g(f(X_{t}))$ for the
inverse function $g$ of $f$, which is strictly increasing, by the same argument as above we know the no arbitrage
property of $f(X_t)$ also implies the no arbitrage property of
$X_t$. \hfill 
\end{proof}
The following example is an application of the above results. This example provides price processes that are not semimartingales and admit an arbitrage within the class 
$S(\mathbb{F})$ but do not admit arbitrage with respect to $S^{0}(\mathbb{F})$.
\begin{example} Let $B_t$ be a Wiener process. Then $-|B_t|$ is a supermartingale. So it satisfies the no arbitrage property with respect to $S^0(\mathbb{F})$ as a result of Proposition~\ref{key lemma}. Then by applying Proposition \ref{monotone1}, we conclude that the processes $X_t=e^{-|B_t|^{\frac{1}{2n+1}}}, n\geq 1$ also have no arbitrage in $S^0(\mathbb{F})$ for any $n\geq 0$. Note that $X_t$ is not a semi-martingale  (see Theorem 71 of \cite{Pro}). \end{example} 

 \bibliographystyle{plain}

\begin{thebibliography}{99}

\bibitem{BS} Bayraktar, E, Sayit, H.: ``Arbitrage free models in markets with transaction costs''. Preprint. Available at http://arxiv.org/abs/0707.0336


\bibitem{DS94} Delbaen, F., Schachermayer, W.: ``A General Version of the
Fundamental Theorem of Asset Pricing'' \emph{\ Math. Ann.} 300, 
463--520, 1994.

\bibitem{DS1994} Delbaen, F., Schachermayer, W.: ``Arbitrage and
  free Lunch with bounded Risk for unbounded continuous Processes.''
\emph{Mathematical Finance}, Vol. 4, 343-348, 1994.


\bibitem{DS95} Delbaen, F., Schachermayer, W.: ``Arbitrage Possibilities in Bessel Processes and their Relations to Local Martingales'' \emph{ Probability Theory and Related Fields} 102, 357--366, 1995.



\bibitem{DH} Delbaen, F., Shirakawa H.: ``No Arbitrage Condition for Positive Diffusion Price Processes", \emph{Financial Engineering and the Japanese Markets}, Vol. 9 (3-4), 159-168, 2002.


\bibitem{KXM} K. D. Elworthy, Xue-Mei Li, M. Yor, '' The importance of strict local martingales; applications to radial Ornstein-Uhlenbeck processes'' Probab. Theory Relat. Fields 115, 325-355, 1999.

\bibitem{Gua} Guasoni, P.:
``No Arbitrage with Transaction Costs, with Fractional Brownian motion
and Beyond'' \emph{Mathematical Finance}, Vol. 16 (2), 469-588, 2006.

\bibitem{HS2000} Heath, D, Schweizer, M..: ``Martingales versus PDEs in Finance: An Equivalence Result with Examples", \emph{Journal of Applied Probability}, Vol. 37,  pp. 947-957,  2000.

\bibitem{HPJ} Jarrow, R., Protter, P., Sayit, H.: 
``No Arbitrage Without Semimartingales'', to appear in \emph{Annals of Applied Probability}. 



\bibitem{KS1991} Karatzas, I., Shreve, S.: \emph{Brownian Motion and Stochastic Calculus}, Second Edition, Springer-Verlag, New York, 1991.

\bibitem{Pro} Protter, P: \emph{Stochastic Integration and Differential 
Equations, Second Edition, Version 2.1}, Springer-Verlag, Heidelberg, 2005.

\bibitem{RY} Revuz, D., Yor, M.: \emph{Continuous Martingales and Brownian Motion}, Third Edition, Springer-Verlag, Heidelberg, 1999.


\bibitem{sin} C.A. Sin,  "`Complications with stochastic volatility models"' Adv. Appl. Prob. 30, 256-268, 1998.


\end{thebibliography}

\end{document}